\documentclass[elsarticle,aps,prl,preprintnumbers,amsmath,amssymb,latexsym,nofootinbib,array,enumerate,letter,twocolumn,superscriptaddress]{revtex4-1}

\usepackage{subfigure}
\usepackage{cancel}
\usepackage{here}
\usepackage{comment,braket}
\usepackage{epsf}
\usepackage{amsmath}
\usepackage{graphics}
\usepackage{amsfonts}
\usepackage{amssymb}
\usepackage{latexsym}
\usepackage{color}
\usepackage{ulem}
\usepackage{adjustbox}
\input{colordvi.tex}
\usepackage{feynmp}

\usepackage[T1]{fontenc} 
\usepackage[english]{babel}
\usepackage[colorlinks=true,linkcolor=blue,urlcolor=blue,citecolor=blue]{hyperref}
\usepackage{amsmath, amssymb, amsfonts}
\usepackage{amssymb,xcolor}
\usepackage{graphicx}
\usepackage{exscale}
\usepackage{graphicx}
\usepackage{amsmath}
\usepackage{latexsym}
\usepackage{mathrsfs}
\usepackage{amsfonts}
\usepackage{amssymb}
\usepackage{float}
\usepackage{braket}
\usepackage{subfigure}
\usepackage{verbatim}
\usepackage{slashed}
\usepackage{indentfirst}
\usepackage{isodateo}
\usepackage{enumerate}
\usepackage[low-sup]{subdepth}
\usepackage{subfigure}
\usepackage{feynmp}
\usepackage{extarrows}

\definecolor{Green}{RGB}{199,238,206}

\usepackage{bm}
\usepackage{dcolumn}
\usepackage{here}
\usepackage{multirow} 
\usepackage{lineno}

\usepackage{isodateo}

\graphicspath{{figs/}}

\newcommand{\be}{\begin{equation}}
\newcommand{\ee}{\end{equation}}
\newcommand{\bea}{\begin{eqnarray}}
\newcommand{\eea}{\end{eqnarray}}

\def\ede{\end{equation}}
\def\bga{\begin{aligned}}
\def\eda{\end{aligned}}
\newcommand{\beq}{\begin{equation}}
\newcommand{\eeq}{\end{equation}}
\newcommand{\bq}{\begin{equation}}
\newcommand{\eq}{\end{equation}}
\newcommand{\ba}{\begin{array}}
\newcommand{\ea}{\end{array}}
\newcommand{\beqa}{\begin{eqnarray}}
\newcommand{\eeqa}{\end{eqnarray}}
\newcommand{\beqs}{\begin{subequations}}
\newcommand{\eeqs}{\end{subequations}}

\def\dis{\displaystyle}

\def\({\left(}
\def\){\right)}

\def\End{\end{document}}
\def\d{\text{d}}
\def\ii{{\tt i}}

\def\be{\beta}

\def\mX{m_X^{}}
\def\vs{\vspace*{1mm}}

\def\hs{\hspace*{0.3mm}}

\def\hsm{\hspace*{-0.3mm}}

\def\hs{\hspace*{0.3mm}}

\def\hsm{\hspace*{-0.3mm}}

\def\vs{\vspace*{1mm}}

\def\End{\end{document}}

\begin{document}



\title{\large
	Search for Light Inelastic Dark Matter\\[1.5mm]
	with Low-Energy Ionization Signatures in PandaX-4T}



\author{Yu-Chen Wang}
\email{wangyc21@sjtu.edu.cn (co-first author).}
\affiliation{State Key Laborotary of Dark Matter Physics, \\
Tsung-Dao~Lee Institute $\&$ School of Physics and Astronomy, \\
Shanghai Jiao Tong University, Shanghai, China}

\author{Youhui Yun}
\email{2461352794@sjtu.edu.cn (co-first author).}
\affiliation{State Key Laboratory of Dark Matter Physics, \\
School of Physics and Astronomy, Shanghai Jiao Tong University, Shanghai, China}

\author{Hong-Jian He} 
\email{hjhe@sjtu.edu.cn (corresponding author).}
\affiliation{State Key Laborotary of Dark Matter Physics, \\
Tsung-Dao~Lee Institute $\&$ School of Physics and Astronomy, \\
Shanghai Jiao Tong University, Shanghai, China}
\affiliation{Department of Physics, Tsinghua University, Beijing, China; \\
\hspace*{-5mm}Center for High Energy Physics, Peking University, Beijing, China}

\author{Yue Meng}
\email{mengyue@sjtu.edu.cn (corresponding author).}
\affiliation{State Key Laboratory of Dark Matter Physics, \\
School of Physics and Astronomy, Shanghai Jiao Tong University, Shanghai, China}
\affiliation{Shanghai Jiao Tong University Sichuan Research Institute, Chengdu, China}
\affiliation{Jinping Deep Underground Frontier Science and Dark\\
Matter Key Laboratory of Sichuan Province, Liangshan, China}

\vspace*{3mm}

\begin{abstract}
Direct detection of light dark matter (DM) is generally difficult due to its small recoil energy.\ 
The inelastic scattering of DM can produce unique signatures 
in the DM direct detection experiments.\  
Using the low-energy unpaired ionization data from PandaX-4T, we newly analyze the probe of   
the exothermic inelastic dark matter (ineDM).\ We demonstrate that PandaX-4T can probe 
the ineDM mass-splitting down to 0.05\,keV and probe the ineDM mass to the sub-GeV range.\  
For the ineDM with a dark photon mediator, we use the PandaX-4T data to impose  
stringent bounds on the mixing parameter between the dark photon and photon. 
\end{abstract}

\maketitle

%
%
%

\section{\hspace*{-2.5mm}Introduction}
\vspace*{-3mm}
\label{sec:intro}
\label{sec:1}

Dark matter (DM) constitutes 85\% of total matter density in the Universe, 
yet its identity and nature remain a great mystery 
because DM has not been found in any detection experiment 
so far\,\cite{Cirelli:2024ssz}.\ 
The DM interactions with the Standard Model (SM) particles (beyond the gravitational interaction) are fully unknown.\  
The current DM searches, especially the direct detection experiments\,\cite{DarkSide:2022dhx,PandaX:2024qfu,LZ:2024zvo,XENON:2025vwd}, 
have put strong constraints on the weakly interacting massive particles (WIMPs).\  
Such DM models consist of a single type of DM particles around the GeV-TeV mass range,  
which scatter elastically off the target nuclei and/or atomic electrons.\  
However, as the direct detection sensitivities start to approach the neutrino floor, 
alternative scenarios such as the axion-like particle (ALP), the dark photon (DP), 
the cosmic-ray boosted DM, and inelastic DM (ineDM) have drawn general  
attentions\,\cite{DarkSide:2022knj,XENON:2022avm,LZ:2023poo,SENSEI:2023zdf,PandaX:2024pme,PandaX:2024sds,XENON:2024znc,LZ:2025iaw}.\  
The new DM scenarios usually consider a different composition of the local DM 
or distinct kinematic effects that change the recoil spectra 
in the DM detection experiments\,\cite{Cirelli:2024ssz}.\

In this Letter, 
we search for exothermic inelastic scattering events of the light ineDM\,\cite{Graham:2010ca,He:2020wjs,He:2024hkr}
in the low-energy ionization data of the PandaX-4T experiment.\ 
In this scenario, the dark matter particle transitions from a heavier state to a lighter state, transferring the mass difference to the final state particles as recoil energy.\
The PandaX-4T experiment employs a dual-phase liquid xenon time projection chamber (TPC) for direct dark matter searches, located in the B2 hall of the China Jinping Underground Laboratory (CJPL-II) \cite{Kang:2010zza,Li:2014rca}.\
It operates by detecting both scintillation and ionization signals generated when incident particles interact with xenon atoms or shell electrons.\ 
These interactions produce prompt scintillation light (the \textit{S}1 signal) and ionization electrons, which drift upward under an electric field and generate delayed electroluminescence (the \textit{S}2 signal) in the gas phase.\ 
The combination of \textit{S}1 and \textit{S}2 signals provides information about the deposited energy, while the time difference between them, together with the spatial pattern of \textit{S}2, enables precise position reconstruction.\ 
The TPC has a cylindrical geometry with a radius of 592 mm and a drift length of 1185 mm between the gate and cathode electrodes, enclosing 3.7 tons of liquid xenon in the active volume.

The detection of the light ineDM requires an extremely low energy threshold, thus the unpaired ionization-only signals (US2) channel is used. In this channel, the analysis threshold is lowered to 0.04 keV electron recoil energy, which is optimized to balance the background suppression and signal acceptance\,\cite{PandaX:2024muv}. The search is performed using Run-0 data (November 28, 2020 – April 16, 2021; 95 calendar days) and Run-1 data (November 16, 2021 - May 15, 2022; 164 calendar days) of the PandaX-4T experiment \cite{PandaX:2024qfu}.

\vspace*{-3mm}
\section{\hspace*{-2.5mm}Exothermic\,Inelastic\,DM\,scattering}
\vspace*{-2mm}
\label{sec:signals}
\label{sec:2}

The exothermic inelastic effect is induced by a pair of near-degenerate DM states $X$ and $X'$, 
with a small mass-splitting $\Delta m \equiv m_{X'}^{}\!-\hsm m_X^{} \!\ll\! m_X^{}$.\ 
In an exothermic inelastic event, $X'\!+\! \rm{SM} \!\to\! X\!+\!\text{SM}\hs$, 
the heavier state $X'$ scatters off a nucleus or an atomic electron and converts to the lighter state $X$.\  
The energy from the mass-splitting is released and transferred to the kinetic energy of the final states.

In this Letter, we consider the light ineDM\,\cite{He:2024hkr} with a mass range around 
$(0.01\!-\!10)$GeV (which spans over three orders of magnitude in mass values).\ 
For most of the parameter space, the nuclear recoil has little detectable signal. 
Instead, the DM-nucleus scattering can contribute to the ionization signal 
through the Migdal effect\,\cite{Migdal,Ibe:2017yqa}.\  
When a nucleus is recoiled, it gains a small velocity relative to its atomic electrons.\  
This may excite or ionize electrons in the outer orbits, and thus triggers ionization signals.\  
Moreover, the DM could directly scatter off the atomic electrons and produce electron recoil events.\  
The differential event rates for the Migdal effect and DM-electron scattering 
are given by\,\cite{He:2024hkr}: 
\begin{widetext}
\beqs 
\vspace*{-4mm}
\begin{align}
\label{eq:rates}
\frac{{\rm d}R_{\rm MIGD}^{}}{{\rm d}E_{\rm ER}^{}\,} & \simeq 
\frac{~\rho_\text{DM}f_{X'}}{\mX}
\frac{\sigma_N^{}}{~2\hs\mu_N^2~}\!\! 
	\int\!\! {\rm d}E_{\rm NR} \d^3 v \frac{\,f(\vec{v}\!+\!\vec{v}_e)\,}{v}
	\sum_{n,\ell} \!
	\frac{~{\rm d}P^c_{q_e}\!(n,\ell\!\to\!E_{e})~}{\,2\hs\pi\hs{\rm{d}}E_{\rm ER}^{}} ,
\\
\label{eq:rates2}
	\frac{\d R_\text{DM-e}^{}}{\d E_{\rm ER}^{}} & \simeq 
	\frac{\rho_\text{DM}f_{X'}}{m_X^{}}
	\frac{\sigma_e^{}}{\,8m_e^2\,} \!\!\int\!\!\d^3 v \frac{\,f(\vec{v}\!+\!\vec{v}_e)\,}{v}
	\!\sum_{n,\ell}\!\frac{1}{\,E_{\rm ER}^{}\!-\!|E_{n\ell}^{}|~}
	\!\int\!\! \d q\, q|f_{n\ell}(E_{\rm ER},q)|^2 ,
\end{align}
\vspace*{-3mm}
\eeqs 
\end{widetext}
where $\rho_\text{DM}$ denotes the total local DM density, 
$f_{X'}$ the fraction of the heavier DM component $X'$, 
$m_X^{}$ the DM particle mass, 
$\sigma_N^{}$ the DM-nucleus scattering cross section at zero momentum transfer, 
$\hs f(\vec{v})\hs$ the local DM velocity distribution function in the galactic frame, 
$\vec{v}_e^{}$ the velocity of the Earth,
$P^c_{q_e}$ the transition probability at the electron momentum transfer $q_e^{}$, 
$\sigma_e^{}$ the DM-electron scattering cross section, 
and $f_{n\ell}^{}$ denotes the ionization probability.\  
Here $E_{\rm{NR}}^{}$ represents the nuclear recoil energy, and $E_{\rm{ER}}^{}$ is the electron recoil energy 
which equals the sum of the kinetic energy $E_e^{}$ of the free electron and the ionization energy $|E_{n\ell}^{}|$
of the $(n,\,\ell)$ orbit.\ Without loss of generality,  we set 
$\rho_{\rm{DM}}^{} \!=\! 0.3\,\rm{GeV}/\rm{cm}^3$ and $f_{X'}^{}\!=\!1\hs$
for the present study.\

Note that although both processes involve the ionization of atomic electrons, 
their probabilities are calculated differently.\ 
This is due to the difference in the momentum transfer: 
for the Migdal effect, the electron momentum transfer (relative to the recoiled nucleus) is
$q_e^{}\!=\! m_e^{} v_{\rm{rel}}^{} \!\sim\! m_e^{} q_{\rm{DM}}^{}/m_N^{} \!\ll\! O(\rm{keV})$, 
but for the DM-electron scattering case, it is $q\hsm\sim\hsm q_{\rm{DM}}^{} \!\gg\! O(\rm{keV})\hs$. 

\vs

The exothermic inelastic effect plays an important role in the event rates, 
as it significantly alters the electron recoil energy $E_{\rm{ER}}^{}$, especially for the light DM.\  
For the DM-nucleus scattering, this effect enhances the overall event rate for Migdal effect 
without much change in the shape of the energy spectrum\,\cite{Bell:2021zkr}.\  
For the DM-electron scattering, this effect results in a peak-like ER spectrum at the mass-splitting, 
i.e.,  $E_{\rm{ER}}^{}\!\approx\!\Delta m\hs$.

\begin{figure}[b]
\vspace*{-6mm}
\centering
\hspace{-6.5mm}
\includegraphics[width=9.2cm]{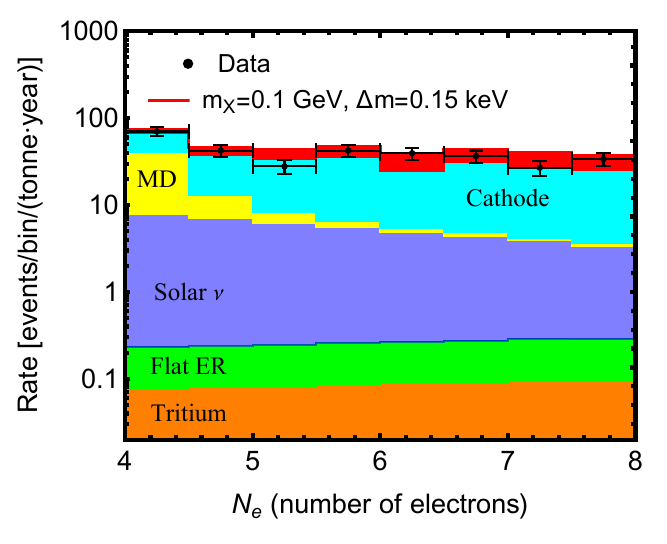}
\vspace*{-7mm}
\caption{\small
US2 event numbers from the PandaX-4T Run-0 and Run-1 combined data,
plotted in S2 (converted to the number of ionized electrons $N_e$).\  
The stacked background components are determined by the background-only fit\,\cite{PandaX:2024muv}.\  
The expected DM signals are shown on top of the backgrounds as the red shadow, 
which consists of contributions from both the Migdal effect and DM-electron scattering.}
\label{fig:bkg}
\label{fig:1}
\vspace*{-4mm}
\end{figure}

\vspace*{-3mm}
\section{Data Selection}
\vspace*{-1.5mm}
\label{sec:exps}
\label{sec:3}

We use the combined US2 data from Run-0 and Run-1 of PandaX-4T \cite{PandaX:2024muv}, 
with effective exposures of 0.49 and 0.55 tonne\,$\cdot$\,year, respectively.\  
We apply four categories of data selection cuts, including S2 pulse-shape, top-to-bottom charge ratio, reconstruction quality, and afterglow veto to suppress micro-discharge (MD) background, cathode background and afterglow effect, 
as described in \cite{PandaX:2024muv}\cite{PandaX:2022aac}.\ 
The region-of-interest (ROI) cut for US2 data requires an \textit{S}2 signal with 
$4\!-\!8$ detected electrons and the absence of an \textit{S}1 signal detected 
by two or more PMTs within the event window, thereby enabling the search 
for exothermic ineDM recoil events with a low energy threshold.\ 
The detection efficiency is extracted from the waveform simulation\,\cite{PandaX:2023sil}, 
which is verified by the $^{220}$Rn, $^{241}$Am-Be and D-D calibration data. \

\begin{figure}[t]
\centering
\hspace*{-8mm}
\includegraphics[width=9.5cm]{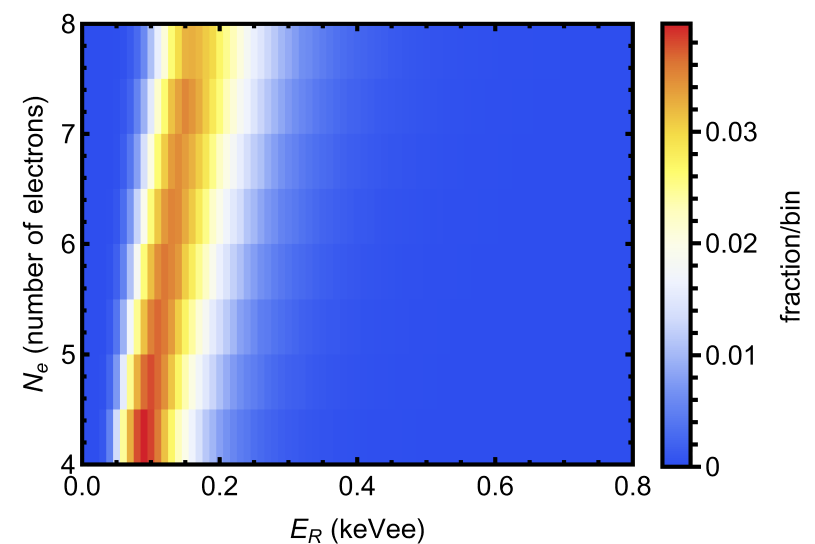}
\vspace*{-5mm}
\caption{\small\hspace*{-1mm}
Response matrix of electron-recoil energy $E_R^{}$ versus the number of ionized electrons $N_e$ in PandaX-4T US2 data.
}
\label{fig:matrix}
\label{fig:2}
\end{figure}

The total number of selected events is 332 within the range of $4\!-\!8$ ionized electrons 
($N_e$) \cite{PandaX:2024muv}.\ These data are shown as the black points in Fig.\,\ref{fig:1}, 
with statistical uncertainties indicated by error bars.\ 
The stacked background components are displayed in Fig.,\ref{fig:bkg}. The background includes flat electron recoil (ER) contributions from $\rm^{214}Pb$, $\rm^{212}Pb$, $\rm^{85}Kr$, and material ER, as well as tritium decay, cathode background, MD noise, and solar neutrinos.\ Their rates are taken from the best-fit results in Ref.\,\cite{PandaX:2024muv} for Run-0 and Run-1.\, 
The dominant backgrounds in the US2 data are (i)\,cathode electrode background, 
estimated using a sideband region of $11\!-\!15$ ionized electrons 
under the assumption of spectral-shape consistency with cathode control samples, 
and (ii)\,MD, whose spectral shape was determined from afterglow-vetoed events and 
whose rate was computed from a sideband region of $2.5\!-\!4$ ionized electrons.

For a more conservative estimate, we adopt the constant-$W$ model\,\cite{Essig:2017kqs}, 
in which the fraction of ionization to total quanta is fixed at $f_e^{} \!=\! 0.83\hs$, 
independent of energy and without recombination effects.\ 
We choose the benchmark parameter inputs of the ineDM model,  
$m_X^{}\!=\!0.1\hs$GeV, $\Delta m \!=\! 0.15\hs$keV, and $\Lambda\!=\!100\hs$GeV.\ 
With these, we compute the DM-proton and DM-electron scattering cross sections as  
$3.0\times 10^{-44}\hs\rm{cm}^2$ and $9.3\times\!10^{-40}\hs\rm{cm}^2$, respectively.\ 

The detection efficiency and the constant-$W$ model are incorporated into the response matrix\,\cite{Supp}, 
which maps the theoretical spectrum to the $N_e$ distribution, as shown in Fig.\,\ref{fig:matrix}.\ 
The matrix is derived from a combination of Run-0 and Run-1 data, weighted by their respective exposures.

\begin{figure}[t]
\centering
\hspace{-7mm}
\includegraphics[height=6.75cm]{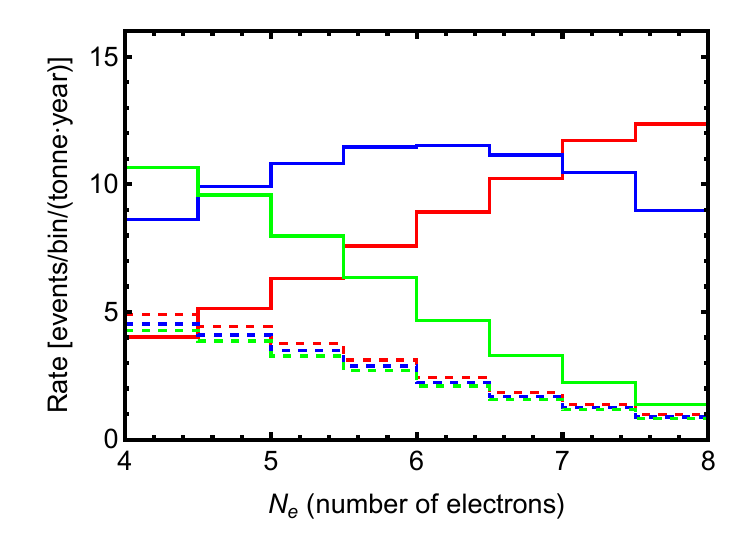}
\vspace*{-8mm}
\caption{\hspace*{-1mm}
\small Expected unpaired S2 event numbers of the exothermic inelastic DM scattering
versus the number of ionized electrons $N_e$.\ 
The (green,\,blue,\,red) curves present the benchmarks with DM mass-splitting 
$\Delta m\hsm\!=\hsm\!(0.08,\hs 0.13,\hs 0.2)\hs${keV}, respectively.\  
The solid curves present the benchmarks with $(m_X^{},\Lambda)\!=\!(0.1,100)$ GeV, 
whereas the dashed curves present the benchmarks with   
$(m_X^{},\Lambda)\!=\!(1,\,250)\hs$GeV.\  
In this plot, the three dashed curves almost coincide. 
}
\label{fig:signals}
\label{fig:3}
\vspace*{1mm}
\end{figure}

\vspace*{-3mm}
\section{Constraints on Dark Photon Mediated Inelastic DM}
\vspace*{-2mm}
\label{sec:limits}
\label{sec:4}

\begin{figure*}[t]
\centering
\includegraphics[height=6.4cm]{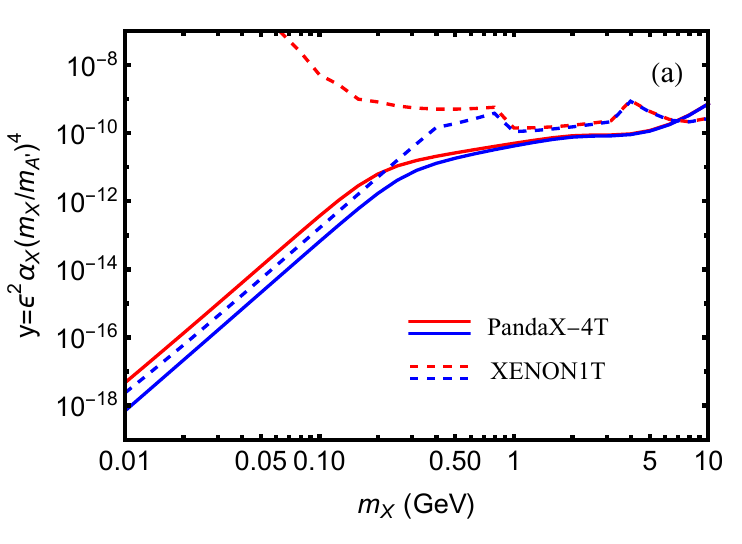}
\hspace*{-3mm}
\includegraphics[height=6.4cm]{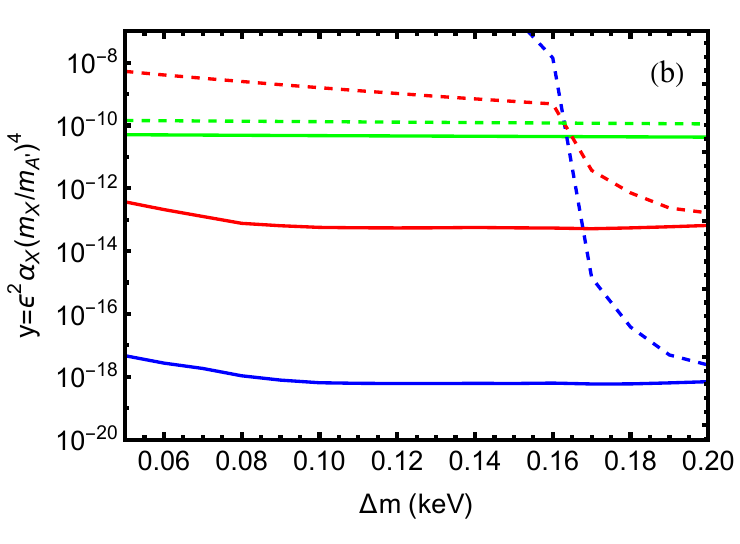}
\vspace*{-6mm}
\caption{\small
Plot-(a): 
Upper bounds (95\%\,C.L.) on the dark sector parameter $y$ versus the inelastic DM mass $m_X^{}$.\  
We derive these bounds from the PandaX-4T US2 data (solid curves) and the XENON1T S2-only data
(dashed curves), respectively.\ 
The (red, blue) curves denote the cases of $\Delta m\!=\!(0.05,\hs 0.2)\,${keV}, respectively.\  
Plot-(b): Upper bounds (95\%\,C.L.) on the dark sector parameter $y$ versus the DM mass-splitting $\Delta m_X^{}$.\  
The solid (dashed) curves are derived from PandaX-4T (XENON1T) data.\  
The (blue,\,red,\,green) curves present the benchmarks with $m_X^{}\!=\!(0.01,\hs 0.1,\hs 1)\hs${GeV} respectively. 
}
\label{fig:limits}
\label{fig:4}
\end{figure*}

In this section, we analyze the the experimental constraints on the inelastic DM (ineDM) 
by using the Run-0 and Run-1 data measured by PandaX-4T experiment\,\cite{PandaX:2024muv}.\  

\vs  

For the ineDM signals, since the detector does not distinguish whether the recoiled electron is produced 
by the Migdal effect or DM-electron scattering, 
we always fit the data with the total event rate, which is the sum of Eq.\eqref{eq:rates} and Eq.\eqref{eq:rates2}.\  
Since the ratio between $\sigma_N^{}$ and $\sigma_e^{}$ is model-dependent, 
the shape of the ER spectrum varies among models.\  
For simplicity, we use a benchmark ineDM model for the present analysis,  
in which a dark photon (DP) serves as a mediator 
between the dark sector and the visible sector\,\cite{He:2024hkr}\cite{Holdom:1985ag}.\  
We describe this benchmark ineDM model in detail based on \cite{He:2024hkr} 
in the Supplemental Material\,\cite{Supp}.\  
The cross sections of the DM-nucleus and DM-electron scattering are given by
\beqs 
\begin{align}
\label{eq:cross-sections}
\sigma_{XN}^{} & \,=\, \dis Z^2 \frac{\,4\hs\alpha g_X^2\epsilon^2\mu^2\,}{~(m_{A'}^2\!+ q^2)^2~}\hs, \\
\label{eq:cross-sections2}
\sigma_{Xe}^{} & \,=\, \dis\frac{\,4\hs\alpha g_X^2\epsilon^2 m_e^2\,}{~(m_{A'}^2\!+ q^2)^2~} \hs, 
\end{align}
\eeqs 
where $Z\!=\!54$ is the charge of the target xenon nucleus, 
$\alpha$ the fine structure constant, 
$g_X^{}$ the DM-DP coupling, 
$\epsilon$ the mixing parameter between the DP and SM photon,
$\mu$ the reduced mass of the DM particle and xenon atom, 
and $m_{A'}^{}$ denotes the DP mass. 
For DP mass $m_{A'}^{}$ much larger than the momentum transfer $q\!\sim\! 10^{-3}m_X^{}$, 
we have $\sigma(q^2)\!\simeq\!\sigma(0)\hs$.\  
Hence, we can simplify the above cross section formulas by absorbing all the dark sector parameters into one variable, 
$y\!\equiv\! g_X^2 \epsilon^2 m_X^4 /(4\pi m_{A'}^4)\hs$.\  
This parameter is connected to the model-independent cutoff scale $\Lambda$ in Ref.\,\cite{He:2024hkr} 
through the relation $\,y\!=\!m_X^4/(4\pi\Lambda^4)\hs$.\  
We plot in Fig.\,\ref{fig:bkg} the expected signal event rates as the red shadow (on top of the background contributions) 
by using the benchmark inputs $m_X^{}\!\!=\!0.1$\,{GeV}, $\Delta m \!=\! 0.15$\,{keV} and $\Lambda\!=\!100$\,{GeV}. 
The red curve stands for the total event rates, including contributions of 
both the backgrounds and DM signals.

We further present a comparison of the expected event numbers 
between various DM benchmarks in Fig.\,\ref{fig:signals}.\  
The solid (dashed) curves describe the case of relatively light (heavy) DM 
with the mass $m_X^{}\!=\!0.1$\,GeV\,($1$\,{GeV}) 
and the cutoff scale $\Lambda\!=\!100$\,{GeV}\,($250\hs${GeV}).\  
The (green,\,blue, red) colors correspond to the mass-splitting 
$\Delta m\!=\!(0.08,$ $0.13, 0.20)\hs${keV}, respectively. 
We see that the solid curves exhibit a peak-like structure, 
which is induced by the DM-electron scattering.\  
This shows that for the DM mass $m_X^{}\!=\!0.1\hs${GeV} (or sub-GeV range in general), 
the DM-electron scattering dominates over the Migdal effect, 
and the shape of the recoil spectrum is highly sensitive to the mass-splitting.\ 
On the other hand, all the three dashed curves almost coincide, 
which indicates that for $m_X^{}\!=\!1$\,GeV (or above) 
the peak-like structures of the DM-electron scattering are no longer significant.\  
Moreover, for GeV-scale DM, 
since the kinetic energy of the incoming DM particle is comparable with or even greater than
the keV-scale mass-splitting, 
the inelastic effect has minor contribution to the rate of Migdal effect.\  
For $m_X^{}\!\gtrsim\!10$\,{GeV}, the dominant contribution to the unpaired S2 signal is from the nuclear recoil.\  
In this case, the inelastic effect becomes negligible.\ 
Hence the inelastic effects are detectable mostly for sub-GeV ineDM.

\vs

For probing such inelastic effect, 
especially the peak-like structure induced by DM-electron scattering, 
we consider light ineDM in the mass range 
$0.01\hs\text{GeV} \!<\! m_X^{} \!<\! 10\hs\text{GeV}$.\ 
The mass-splitting is around $0.05\hs\text{keV} \!\!<\! \Delta m \!<\! 0.20\hs\text{keV}$, 
for which the data between $(4\!-\!8)N_e$ can provide the most sensitive probe 
(as shown in Fig.\,\ref{fig:2}).\  
Given the unpaired S2 data and the background model, 
we derive the upper bounds (95\%\,C.L.) on $y$ by setting the test statistics,  
$\chi^2\!=\!\chi_{\min}^2\hsm +\hsm 1.64^2$, 
where only statistical errors of the data are taken into account.\  
We present these bounds as the solid curves in Fig.\,\ref{fig:limits}.\  
The plot\,(a) presents the constraints on the cases of the mass-splitting 
$\Delta m\!=\!(0.05, 0.2)\,$keV as the (red,\,blue) solid curves, respectively.\  
In plot\,(b), the (blue,\,red,\,green) solid curves present 
the bounds on the cases of $m_X^{}\!\!=\!(0.01,\hs 0.1,\hs 1)\hs${GeV}, respectively.\  
For comparison, we have further derived the constraints using 
the available XENON1T S2-only data\,\cite{XENON:2019gfn}, 
shown as the dashed curves in each plot.\  
Their colors correspond to the same benchmark parameters as for the solid curves. 

\vs 

From Fig.\,\ref{fig:limits}(a), we see that the sensitivity bound on
the ineDM parameter $y$ scales roughly 
as $m_X^5$ for $m_X^{}\!\lesssim\!0.3\hs$GeV (except for the red dashed curve).\  
This is due to the fact that the local DM number density is proportional to $1/m_X^{}$ 
and $y$ itself scales as $m_X^4$.\  
Other dependence on $m_X$ is negligible in this case, 
because below the GeV scale and for $\Delta m\!\lesssim\! O(\rm{keV})$, 
the DM-electron scattering dominates over the Migdal effect (as discussed for Fig.\,\ref{fig:signals});
this means that the total rates depend mainly on the mass-splitting $\Delta m$ rather than the DM mass $m_X^{}\hs$.\  
On the other hand, Fig.\,\ref{fig:limits}(b) shows that the sensitivity bound varies 
with the mass-splitting $\Delta m\hs$.\  
Since the exothermic inelastic DM-electron scattering produces a sharp peak-like structure 
in the ER spectrum at $E_{\rm{ER}}^{}\!\simeq\! \Delta m\hs$, 
the detection threshold of ER energy determines the threshold of $\Delta m\hs$.\  
We see in Fig.\,\ref{fig:limits}(b) that for $m_X^{}\!\!<\!1\hs${GeV}, 
the sensitivity bound of XENON1T weakens rapidly as $\Delta m$ drops below 
its ER threshold 0.186\,keV \cite{XENON:2019gfn}.\ 

In summary, for the sub-GeV light DM, the exothermic inelastic DM-electron scattering 
can be the dominant process in the direct detection experiments.\  
With a lower energy threshold for ionization signal detection, 
the sensitivity of the sub-GeV DM search can be largely enhanced.

\vspace*{-6mm}
\section{Conclusions}
\vspace*{-3mm}
\label{sec:conclusion}
\label{sec:5}

In this Letter, we newly analyze the detection of the exothermic inelastic DM (ineDM) signals, 
by using the low-energy unpaired ionization data of the PandaX-4T experiment.\  
We apply this analysis to the exothermic ineDM with a dark photon mediator, 
and derive stringent new upper bounds 
on the dark sector parameter space of $m_X^{}$ versus $y\hs$  
or $\Delta m$ versus $y\hs$ as presented by the solid curves in Fig.\,\ref{fig:4}(a)-(b),
which are stronger than the bounds (dashed curves) derived from the available XENON1T S2-only data.\ 
Our study demonstrates that the exothermic inelastic scattering channel provides sensitive probes  
of the parameter space of the light ineDM (including the photon-dark photon mixing parameter), 
especially in the sub-GeV mass range.

\vspace*{5mm}
\noindent
{\bf\large Acknowledgements} 
\\[1mm]
We thank PandaX collaboration for cooperation and support 
during the course of this project.\ 
The works of HJH and YCW were support in part
by the National Natural Science Foundation of China (NSFC)
Grants 12175136 and 12435005.\ 
The works of YM and YY were supported in part by the Ministry of Science and Technology of China (No.\,2023YFA1606204), the National Natural Science Foundation of China (No.\,12222505) and Sichuan Science and Technology Program (No. 2024YFHZ0006).
This research was also supported in part 
by the State Key Laboratory of Dark Matter Physics, 
by the Key Laboratory for Particle Astrophysics and Cosmology (MOE), 
and by the Shanghai Key Laboratory for Particle Physics and Cosmology.\ 

\noindent 
In this work, YCW contributed to the theoretical spectrum calculations and the analysis of constraints on ineDM, while YY contributed to the experimental data analysis.\ All authors participated in the editing process and approved the final version of the manuscript.


\baselineskip 17pt

\vspace{5mm}
%

\newpage
\onecolumngrid

\renewcommand{\thesection}{S\arabic{section}}
\renewcommand{\theequation}{S\arabic{equation}}
\renewcommand{\thefigure}{S\arabic{figure}}
\renewcommand{\thetable}{S\arabic{table}}
\setcounter{equation}{0}

\newpage 
\section*{Supplemental Materials}
\vspace*{1.5mm}

\subsection{A.~Response Matrix of PandaX-4T}
\label{app:response}
\label{app:A}
\vspace*{1.5mm}

The response matrix (shown in Fig.\ref{fig:matrix}) is stored in a CSV file, where each bin represents the efficiency of converting a specific electron recoil energy into the number of detectable electrons using the constant-W model\,\cite{Essig:2017kqs, PandaX:2024med}.\  This matrix encodes the full detection efficiency and is obtained as the exposure-weighted average of Run-0 and Run-1.

\subsection{B.~Inelastic DM with Dark Photon Mediator}
\label{app:model}
\label{app:B}
\vspace*{1.5mm}

The existing bounds from all experiments and observations suggest that
the interactions between DM and the SM particles are very weak.\  
The dark photon provides a natural mediator connecting the dark sector to the visible world 
through a tiny kinematic mixing between a dark $U(1)$ gauge field (the dark photon) and the SM photon: 
\beq
\mathcal{L}_{\rm{mix}}^{} \supset \frac{\,\kappa\,}{2}\bar{A}'_{\mu\nu}\bar{B}^{\mu\nu} , 
\eeq
where $\bar{A}'_{\mu\nu}\!=\partial_\mu\bar{A}'_\nu\!-\partial_\nu\bar{A}'_\mu$ is the dark $U(1)_X^{}$ field strength, 
and $\bar{B}_{\mu\nu}^{}\!=\hsm\partial_\mu\bar{B}_\nu\!-\hsm\partial_\nu\bar{B}_\mu$ the SM $U(1)_Y^{}$ field strength. 

For the inelastic DM (ineDM)\,\cite{He:2024hkr}, 
we consider a complex scalar DM, $\hat{X}\!=\!(X\! +\hsm\ii X')/\hsm\sqrt{2\,}$, 
having charge $+1$ under the dark $U(1)_X^{}$ gauge group :
\begin{align}
\mathcal{L}_\text{DM} \supset &\, 
(D_\mu \hat{X})^\dagger (D^\mu \hat{X}) - m_X^2 |\hat{X}|^2 + \delta^2 (\hat{X}^2\!+\hsm\rm{h.c.}) 
\nonumber \\
= &\, 
\frac{1}{\,2\,}\partial_\mu X \partial^\mu X \!+\! \frac{1}{\,2\,}\partial_\mu X' \partial^\mu X' 
\!-\!\frac{1}{\,2\,}(m_X^2\!-\!2\delta^2)X^2 -\frac{1}{\,2\,}(m_X^2\!+\!2\delta^2)X'^2 
\nonumber \\
&\, + g_X^{} (X'\partial^\mu X \!-\! X\partial^\mu X') \bar{A}'_\mu\!+\!\frac{1}{\,2\,}g_X^2\bar{A}'^2(X^2\!+\!X'^2)\,,
\end{align}
where $g_X^{}$ is the coupling constant of the dark $U_X^{}(1)$ gauge field.\  
For $\delta^2\!\ll\! m_X^2$, this gives rise to a pair of near-degenerate DM states $X$ and $X'$,
whose mass eigenvalues are given by $m_X^{}(1\!\mp\!\delta^2/m_X^2)$,  
with a small mass-splitting $\Delta m \!\simeq\! 2\delta^2/m_X^{}\,$. 

After the spontaneous symmetry breaking in both the dark sector and visible sector, 
we diagonalize the kinetic terms and mass terms and obtain the dark photon portal interactions
at leading order: 
\beq
\mathcal{L}_{\rm{int}}^{} \,\supset\,  
g_X^{} A'_\mu J_X^\mu - e \hs\kappa\hs \cos\theta_W^{} A'_\mu J_\text{em}^\mu \,,
\eeq 
where $A'_\mu$ is the physical dark photon field (in mass eigenstate), 
$J_X^{\mu}\!=\!(X'\partial^\mu X \hsm -\hsm  X\partial^\mu X')$ the dark current, 
$\theta_W$ the weak mixing angle,  
and $J_\text{em}^\mu$ the electromagnetic current.\  
These vertices generate the cross sections of the DM-nucleus and DM-electron scattering
as shown in Eq.\eqref{eq:cross-sections} and Eq.\eqref{eq:cross-sections2}. 

The kinematic mixing also leads to the following interactions\,\cite{He:2024hkr}: 
\begin{align}
\mathcal{L}_\text{int}^{} ~\supset~ & 
	g_X^{}\kappa\hs s_W^{}\hsm  Z_\mu\hsm J_X^\mu 
     - \frac{\,g\hs m_{\bar{A}'}^2\,}{\,c_W^{}M_{\bar Z}^2\,}\kappa\hs s_W^{}A'_\mu\!\! J_Z^\mu\,.
\end{align}
where $J_Z^\mu$ is the weak neutral current.\ 
For $\Delta m \!<\hsm 2m_e^{}$, the dominant channel of $X'$ decay is $\,X'\!\!\to\! X\nu\bar\nu\,$,
which is induced by the above interactions.\ 
The decay width is given by 
\begin{align}
	\Gamma_{X'\rightarrow X\nu\bar\nu}^{} 
    \,\approx\,  \left(4\!\times\! 10^{43}\text{yrs}\right)^{\!-1}\!
	\(\!\!\frac{\Lambda}{\,100\hs\text{GeV}\,}\!\)^{\!\!-4}\!\!
	\(\!\frac{\Delta m}{\,1\hs\text{keV}\,}\!\)^{\!\!9} \,.
\end{align}
This shows that the lifetime of $X'$ is much longer than the age of the Universe. 
Hence it will remain abundant today once created in the early universe.


\end{document}